# Electronic and optical properties of Ti$_{1-x}$Cd$_x$O$_2$: A first-principles prediction

X.D. Zhang[1]

*Department of Health Physics, Institute of Radiation Medicine, Chinese Academy of Medical Sciences and Peking Union Medical College, Tianjin 300192, People's Republic of China*

M.L. Guo

*Department of Fundamental Subject, Tianjin Institute of Urban Construction, Tianjin 300384, People's Republic of China*

C.L. Liu and W.X. Li

*Department of Applied Physics, School of Science, Tianjin University, Tianjin 300072, People's Republic of China*

X.F. Hong

*National Key Laboratory for Nuclear Fuel and Materials, Nuclear Power Institute of China (NPIC), Chengdu, 610041, People's Republic of China.*

A first-principles study has been carried out to evaluate the electronic and optical properties of rutile Ti$_{1-x}$Cd$_x$O$_2$ as a possible photocatalytic material. It was found that Cd incorporation lead to the enhancement of *p* states in the top of valence band and the decrease of band gap. The optical transition between Cd *p* and O *p* enhances gradually and shifts to high energy range with increasing Cd concentration. Furthermore, optical absorption of Ti$_{1-x}$Cd$_x$O$_2$ increases in the visible range.

[1] Corresponding author (e-mail: xiaodongzhang@irm-cams.ac.cn).
Tel:+86-22-85682375

Titanium dioxide ($TiO_2$) is an effective photocatalyst for the remediation of organic pollutants and promising for the possible application to the solar energy conversion.[1] Unfortunately, due to the wide band gap, it can absorb only ultraviolet light, which limits its some possible applications in photocatalyst. Therefore, for the sake of utilizing the wider range of solar light, it is necessary to enhance the optical absorption in the visible range.

It has been demonstrated that this problem can be solved by the appropriate ion dopants. Especially, N-doped $TiO_2$ is the most promising to enhance the optical absorption in the visible range and solves this bottleneck.[2,3,4] However, due to strongly localized N $p$ states in the top of valence band,[5] photocatalytic efficiency of N-doped $TiO_2$ decreases in the high N concentration.[6] The optimal N concentration for photocatalyst is only around 2%,[6,7] which is not as positive as the previous expectations.[2,3,4] In addition, the N doping may induce the some O vacancies, and its thermal stability has also been challenged.[8] Transition metal doping can promote the photocatalytic property.[9] However, it is still controversial. Some metal ions, such as Nb, Mn, may produce not only shallow traps but also deep-level defects which act as recombination centers.[10,11] Therefore, it is essential to attempt the other metal ions doping for overcoming their disadvantages.

CdO is a compound semiconductor with direct band gap of 2.3 eV, and Cd incorporation into metal-oxide-semiconductor can modulate the band gap effectively.[12,13] Pure $Ti^{+4}O^{-2}_2$ is a wide band-gap semiconductor with the O-$p$ band filled and the Ti-$d$ band empty. When a Cd atom replaces a Ti atom the resulting system is metallic because of the lack of two electrons necessary to fill up the O-$p$ band.[14] Thus, it can be expected that Cd incorporation in $TiO_2$ can result in some unusual optical transitions and the optical absorption in the visible range. In this letter, electronic and optical properties of rutile $Ti_{1-x}Cd_xO_2$ have been investigated by using the first-principles in order to reveal the mechanism of optical transitions and absorption.

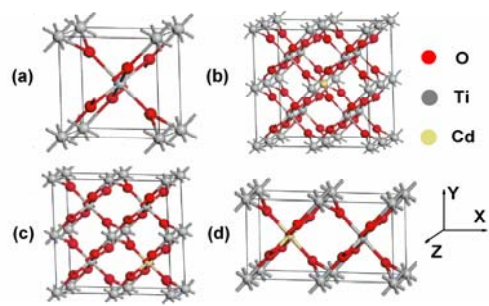

Fig.1 The crystal structure of $Ti_{1-x}Cd_xO_2$ when (a) x=0; (b) x=0.0625; (c) x=0.125; and (d) x=0.25.

The calculations were performed with CASTEP code, based on density functional theory (DFT) using a plane-wave pseudopotential method.[15] We used the generalized gradient approximation (GGA) in the scheme of Perdew-Burke-Ernzerhof (PBE) to describe the exchange-correlation functional.[16] The norm-conserving pseudopotential was used with $2s^22p^4$ and $3d^24s^2$ as the valence-electron configuration for the oxygen and titanium atoms, respectively, to describe the electron-ion interaction.[17] Rutile $Ti_{1-x}Cd_xO_2$ supercell contained 48, 24, 12 and 6 atoms, which corresponded to x=0.0625, 0.125, 0.25 and 0, respectively (see Fig.1). Each supercell contained only one Cd atom in order to avoid the interaction of doping atoms. We chose the energy cutoff to be 600 eV, and the Brillouin-zone sampling mesh parameters for the $k$-point set were 2×2×3, 2×2×7 and 2×4×7 for x=0.0625, 0.125 and 0.25, respectively.[18] The charge densities were converged to $2 \times 10^{-6}$ eV/atom in the self-consistent calculation. In the optimization process, the energy change, maximum force, maximum stress and maximum displacement tolerances were set as $2\times10^{-5}$ eV/atom, 0.05 eV/Å, 0.1 GPa and 0.002 Å, respectively. The scissors operation of 0.8 eV has been carried out in optical absorption of $Ti_{1-x}Cd_xO_2$ system.[4,7,12]

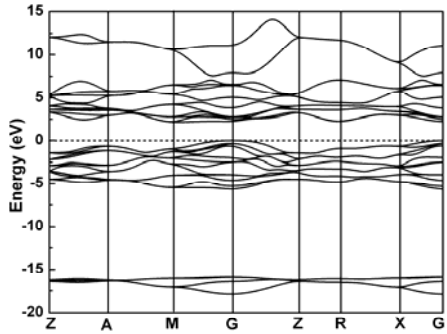

Fig.2 Band structure of rutile $TiO_2$

Fig.2 shows the band structure of pure $TiO_2$. The band gap is about 2.21 eV at highly symmetric G point, which is less than the experimental value of 3.0 eV.[19] The underestimated band gap can be due to the choice of exchange-correlation energy. The valence band mainly consists of the 2p, 2s states of O and 3d states of Ti. In the uppermost valence band, O 2p states are predominantly found between -6 and 0 eV, while O 2s states appear in the range from -18 to -15.5 eV. Ti 3d states give rise to some bands in the energy range from -6 to -3 eV. The lowest conduction band is dominated by Ti 3d states. The calculated electronic structures described in this work are consistent with the previous results.[4,7,14,20]

Fig.3 gives the partial density of states (PDOS) of $Ti_{1-x}Cd_xO_2$. Compared with the pure $TiO_2$ in Fig.3 (a), the Cd 4d states in $Ti_{1-x}Cd_xO_2$ have been observed in the energy range from -9.5 to -7.5 eV[Fig.3(b)-(e)]. Meanwhile, the Cd 4d states enhance gradually with the increasing Cd concentrations, while the Ti 3d states between -6 and -3 eV decrease slightly. O 2s states have slightly shifted to low energy range. Cd incorporation induces the increasing p states in the top of valence band, and thus band gap decreases, which is very similar with the previous calculated result.[14] The band gaps of $Ti_{1-x}Cd_xO_2$ are 2.0 eV, 1.8eV and 1.6eV, which correspond to x=0.0625, 0.125 and 0.25, respectively. In addition, in high concentration of x=0.25, the slight blueshift in conduction band are observed. All these variations arouse our great interesting to the optical properties of $Ti_{1-x}Cd_xO_2$.

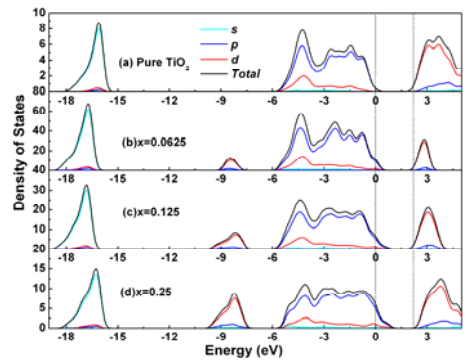

Fig.3 The PDOS of $Ti_{1-x}Cd_xO_2$ when (a) x=0; (b) x=0.0625; (c) x=0.125; and (d) x=0.25.

The interaction of a photon with the electrons in the system can be described in terms of time-dependent perturbations of the ground-state electronic states. Optical transitions between occupied and unoccupied states are caused by the electric field of the photon. The imaginary part of the dielectric function $\varepsilon_2(\omega)$ can be calculated from the momentum matrix elements between the occupied and unoccupied wave functions and the real part $\varepsilon_1(\omega)$ of the dielectric function can be evaluated from the imaginary part $\varepsilon_2(\omega)$ by the famous Kramer-Kronig relationship.[12,16,21] Fig.4 exhibits the imaginary part of dielectric function $\varepsilon_2(\omega)$ of $Ti_{1-x}Cd_xO_2$. To the pure $TiO_2$ [Fig.4 (a)], the optical transition of 3.71 eV ($E_2$) was only observed, which is ascribed to the intrinsic transition between O 2*p* in the highest valence band and Ti 3*d* in the lowest conduction band. After Cd incorporation, optical transitions ($E_1$) are noticed, which is ascribed to the transitions between O 2*p* and Cd *p*. This transition ($E_1$) enhanced gradually and the corresponding $\varepsilon_2(\omega)$ has increased sharply from 3.0 to 10.4. In particular, the $\varepsilon_2(\omega)$ of $E_1$ is even stronger than that of $E_2$ in x=0.25, which indicates that $E_1$ is stronger than intrinsic optical transition in this configuration. Meanwhile, with the increasing Cd concentration, the value of $E_1$ has increased from 0.5 eV to 1.0 eV due to increasing Cd *p* states in the top of valence band and is consistent with the results of the PDOS. The values of $E_2$ are 3.49, 3.73 and 3.95 eV, which correspond to Cd concentration of x=0.0625, 0.125 and 0.25, respectively. The increasing $E_2$ value indicates the increase of the intrinsic transition. Cd incorporation into $TiO_2$ induces the slight shift of conduction band, and furthermore influences the intrinsic transition of $E_2$, which is obvious in the systems with high lying *d* bands.[22] It is clear in inset that $E_2$ is the intrinsic transition between valence band and conduction band, while $E_1$ is due to transition between valence band and Cd *p* states. This result indicates that optical transition between Cd *p* states and conduction band can't take place, and thus band gap narrowing has not induced the appearance of the corresponding transition.

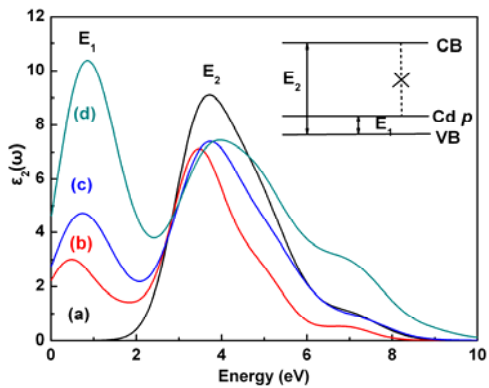

Fig.4 The imaginary part of dielectric function $\varepsilon_2(\omega)$ of $Ti_{1-x}Cd_xO_2$ when (a) x=0; (b) x=0.0625; (c) x=0.125; and (d) x=0.25. $E_1$ and $E_2$ is the O 2*p*-Cd *p* and O 2*p*-Ti *d* transitions, respectively. Inset is the outline of optical transitions in $Ti_{1-x}Cd_xO_2$.

Fig.5 shows the optical absorption spectra of $Ti_{1-x}Cd_xO_2$ under 0.8 eV scissors operation.[4,7,12] It is clearly observed that the Cd incorporation induces the increasing optical absorption in visible range of 380-780 nm. The absorption peak has shifted to short wavelength range (high energy range), which is in good agreement with $\varepsilon_2(\omega)$ of $E_1$ in Fig.4. It indicates that the contribution of the visible absorption is originated from $E_1$ while not band gap narrowing. Optical absorption band edges shift slightly to the short wavelength range in the configuration of x=0.0625 and have no obvious shift in the other configurations. It is noting that the lower Cd concentration of x=0.0625 results in the lower optical absorption due to the weak Cd *p* states in the top of valence band. Therefore, we can expect that lower Cd concentration in $TiO_2$ is difficult to achieve the photocatalyst. The calculated results suggest that the $Ti_{1-x}Cd_xO_2$ with higher concentration is promising as an alternative photocatalytic material.

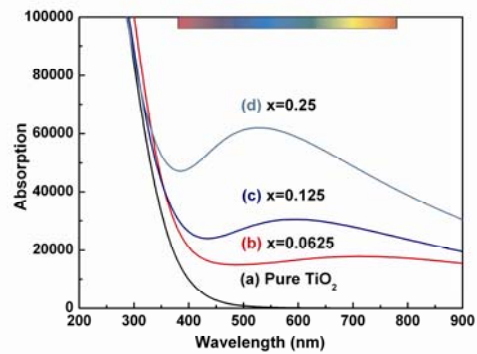

Fig.5 Optical absorption spectra of $Ti_{1-x}Cd_xO_2$ when (a) =0; (b) x=0.0625; (c) x=0.125; and (d) x=0.25.

In summary, we carrioud out a first-principles study to evaluate the electronic and optical properties of rutile $Ti_{1-x}Cd_xO_2$. Cd incorporation leads to the enhancement of *p* states in the top of valence band and the decrease of band gap. However, the band gap narrowing has no contribution to the enhancement of optical absorption in the visible range. The optical transition between valence band and Cd *p* states enhances. As a result, the optical absorption of $Ti_{1-x}Cd_xO_2$ enhances in the visible range. The results indicate that the Cd incorporation $TiO_2$ is very promising for the modulating the visible photoresponse and as a photocatalytic material.

Authors would like to thank Professor Mike Payne (University of Cambridge) for some helpful discussions. This work is supported by the nuclear materials project (Grant. No. A0120060584). Professor C.L.Liu would like to thank the support of the National Natural Science Foundation of China (Grant No.10675089) and Natural Science Foundation of Tianjin (Grant No. 06YFJMJC01100).